\documentclass{article}

\usepackage[final]{aaj2026}
\usepackage[numbers]{natbib}

\usepackage[utf8]{inputenc}
\usepackage[T1]{fontenc}
\usepackage{hyperref}
\usepackage{url}
\usepackage{booktabs}
\usepackage{amsfonts}
\usepackage{nicefrac}
\usepackage{microtype}
\usepackage{xcolor}
\usepackage{xspace}
\usepackage{enumitem}

\title{Building an Internal Coding Agent at Zup: \\ Lessons and Open Questions}

\author{%
  Gustavo Pinto \\
  Zup Innovation\\
  Brazil \\
  \texttt{gustavo.pinto@zup.com.br} \\
  \And
  Pedro Eduardo de Paula Naves \\
  Zup Innovation\\
  Brazil \\
  \texttt{pedro.paula@zup.com.br} \\
  \And
  Ana Paula Camargo \\
  Zup Innovation\\
  Brazil \\
  \texttt{ana.camargo@zup.com.br} \\
  \And
  Marselle Silva \\
  Zup Innovation\\
  Brazil \\
  \texttt{marselle.silva@zup.com.br} \\
}

\begin{document}

\newcommand{\gnote}[1]{\textcolor{red}{[[[#1]]]}}
\newcommand{\codegen}{\textsf{CodeGen}\xspace}
\def\xxx{{\huge XXX}}

\maketitle

\begin{abstract}
Enterprise teams building internal coding agents face a gap between prototype performance and production readiness. The root cause is that technical model quality alone is insufficient---tool design, safety enforcement, state management, and human trust calibration are equally decisive, yet underreported in the literature. We present \codegen, an internal coding agent at Zup, and show that targeted tool design (e.g., string-replacement edits over full-file rewrites) and layered safety guardrails improved agent reliability more than prompt engineering, while progressive human oversight modes drove organic adoption without mandating trust. These findings suggest that the engineering decisions surrounding the model---not the model itself---determine whether a coding agent delivers real value in practice.
\end{abstract}

\section{Introduction}

Large language models (LLMs) are increasingly being integrated into software development workflows through coding agents---systems that go beyond token-level code completion to autonomously read files, edit source code, execute shell commands, and iterate on their own outputs~\citep{wang2024survey, fan2023large}. Tools such as GitHub Copilot~\citep{peng2023impact}, Cursor, and Claude Code have demonstrated that LLM-based agents can accelerate routine development tasks, and a growing number of organizations are now exploring how to build or customize such agents for internal use~\citep{rasheed2024codepori}.

However, building a coding agent that performs well on benchmarks or in isolated demonstrations is fundamentally different from deploying one that developers actually use in their daily work. In enterprise settings, agents must operate on real codebases with complex build systems, interact with internal tooling and CI/CD pipelines, and earn the trust of developers who are accountable for the code the agent produces. For example, an agent that rewrites entire files risks introducing subtle truncation errors in large codebases, while an agent with unrestricted shell access can execute destructive commands such as \texttt{rm~-rf} or unauthorized \texttt{git~push~-{}-force} on a developer's machine. These are not hypothetical risks---they are failure modes that emerged repeatedly during the development of the system described in this paper\footnote{\url{https://github.com/openai/codex/issues/3934}}.

When these engineering challenges are not addressed, there are consequences such as: 1) agents produce unreliable edits that developers must manually verify and correct, 2) safety incidents erode trust and stall adoption, and 3) teams invest months building prototypes that never transition to production use. In a landscape where multiple organizations are simultaneously building internal coding agents, the absence of shared knowledge about these practical challenges means that engineering teams are likely rediscovering the same failure modes independently---wasting effort that could be directed toward advancing the state of the practice.

Existing literature on LLM-based coding agents has focused primarily on model-level performance---benchmark evaluations~\citep{jimenez2024swebench, xu2022systematic}, prompt engineering techniques, and reasoning strategies~\citep{yao2023react, yao2023tree}. Little has been reported about the engineering decisions that determine whether a coding agent succeeds in production: how tools should be designed and specified for LLM consumption~\citep{schick2023toolformer, qin2024toolllm}, how safety policies should be enforced across tools with overlapping capabilities, how session state should be managed for resilience, and how human oversight should be structured to enable adoption without mandating blind trust~\citep{parasuraman2000model}. These dimensions are not secondary concerns; in our experience, they proved more consequential than model selection or prompt optimization for the system's real-world effectiveness.

This paper addresses this gap by presenting the key architectural and engineering decisions behind \codegen, an internal coding agent developed and deployed at Zup Innovation\footnote{\url{https://www.zup.com.br}}, for use by enterprise software development teams. We focus on four core dimensions that shaped the system's evolution: model strategy (how reasoning is delegated to the LLM), framework strategy (when to build manually vs.\ adopt orchestration frameworks), tooling strategy (how tools are designed, specified, and constrained), and human oversight strategy (how trust is calibrated during adoption).


Among the key findings: (1)~refining tool descriptions, parameter schemas, and error contracts produced more consistent improvements in agent reliability than prompt engineering alone; (2)~safety guardrails must be enforced holistically across the entire tool manifest, since restricting one tool is ineffective when equivalent capabilities are accessible through another; (3)~a progressive human oversight model---where developers begin in approval mode and organically migrate to autonomous mode---proved critical for enterprise adoption; and (4)~implementing the agentic loop manually before adopting frameworks gave the team the understanding necessary to evaluate later abstractions on their merits, a strategy validated when modern frameworks converged on the same design.

The contributions of this paper are:
\begin{itemize}[nosep]
  \item A description of \codegen's architecture---a three-tier system comprising a CLI executor, a FastAPI backend, and a centralized orchestration engine (Maestro)---including its state management, resilience mechanisms, and deployment pipeline.
  \item A catalog of 13 concrete design decisions across framework selection, tool design, safety enforcement, and human oversight, each accompanied by the trade-off that motivated it and the lesson derived from it.
  \item An analysis of how tool specification quality, cross-tool safety policy consistency, and progressive trust calibration affect agent reliability and adoption in enterprise environments.
  \item Six open questions on tool manifest design, reasoning boundaries, cross-tool safety enforcement, adaptive trust models, long-term agent memory, and quality assurance for agent-generated code.
\end{itemize}

\section{Origins of CodeGen}

In early 2024, enterprise AI coding tools were shifting from passive assistants to autonomous agents. Products such as Devin and Cursor attracted significant attention for their ability to not only suggest code but also execute multi-step tasks autonomously---reading files, running commands, and iterating on their own output. For Zup, this shift had immediate strategic implications. The company's flagship developer platform, StackSpot AI~\cite{stackspotai:pinto}, offered IDE extensions that operated as conversational assistants: developers could ask questions and receive code suggestions, but the extensions could not take actions in the environment.

An initial assessment revealed that adding agentic capabilities---file editing, command execution, iterative tool-calling loops---to StackSpot AI's enterprise codebase would require substantial changes across the IDE extensions, the backend, and the model integration layer. The platform's roadmap, oriented toward a broader agent infrastructure rather than code generation specifically, could not absorb this scope. Two members of the development team decided to build a proof of concept independently. The goal was to construct a minimal but functional agentic system---a CLI-based coding agent backed by a lightweight orchestration backend---and demonstrate that the agentic pattern could work within Zup's infrastructure. The POC was deliberately scoped outside the StackSpot AI codebase to avoid dependency on its release cycle. This independence freed agentic from the platform's backlog constraints, the team could iterate rapidly, make opinionated design choices, and validate ideas with real developer usage in days rather than quarters. What began as a side project gradually evolved into \codegen, a production ready agentic-system used every day by real developers. 

\section{CodeGen Internals}

\codegen is an internal coding agent developed at Zup that assists software developers by autonomously performing code-related tasks such as editing source files and executing shell commands. Unlike token-level code-completion tools~\citep{chen2021evaluating}, \codegen operates as a fully agentic system grounded in the ReAct (Reasoning + Acting) paradigm~\citep{yao2023react}: given a natural-language prompt, it reasons about the task, selects and invokes tools, observes their outputs, and iterates until the task is complete. The system delegates internal reasoning to a LLM while providing the structural scaffold---context assembly, iterative tool dispatch, and output re-injection---that enables the model to act over successive observations. A configurable thinking effort parameter controls reasoning depth, and a dedicated planning mode was introduced to allow developers to review a proposed action sequence before execution begins.



\subsection{Architecture}\label{sec:architecture}

\codegen's architecture comprises three components,  organized as follow: a CLI interface, a backend API, and a maestro component. 

The first component is the \textbf{CLI} (Node.js client), responsible for user interaction and local tool execution. 
\codegen was designed as a CLI primarily for portability: since all major IDEs---VSCode, IntelliJ, Visual Studio, among others---already include built-in terminal emulators, a single CLI serves as a unified interface across the diverse set of development environments used within the enterprise. Building native IDE extensions would have required developing and maintaining a separate plugin for each IDE, multiplying engineering effort without proportional benefit. A CLI also enabled tighter integration with the enterprise's internal tools and workflows, offering a level of customization and execution control that off-the-shelf, general-purpose coding agents could not match.

A CLI also enabled tighter integration with the enterprise's internal tools and workflows, offering a level of customization and execution control that off-the-shelf, general-purpose coding agents could not match. Although the CLI is the primary interface, the architecture supports multiple client types---including IDE plugins and VM-based executors---through the same communication protocol. 

The second component is the \textbf{Backend API}, built with FastAPI and responsible for maintaining connectivity with clients, handling authentication, and routing requests. The backend exposes a REST API for task lifecycle management (task creation, log retrieval, user-initiated actions) and supports two real-time communication channels: WebSocket connections for executor clients (CLI, IDE plugins) that require bidirectional tool dispatch, and Server-Sent Events (SSE) for web portal interfaces that consume task updates as a read-only stream. This dual-channel design allows different client types to interact with the system through the protocol best suited to their needs.

The third component is the \textbf{Maestro}, the orchestration engine that controls the agentic loop. Upon initialization, a bootstrap phase collects environment metadata---operating system, working directory, recent git history, project structure, etc. When the user issues a prompt, the backend sends the LLM a payload containing the system prompt, conversation history, bootstrap metadata, and a tool manifest describing all available tools and their parameter schemas. The LLM either responds with text or requests a tool call; in the latter case, the Maestro relays the invocation to the CLI via WebSocket, collects the result, and feeds it back to the model. This cycle repeats until the model emits a final textual response instead of a tool call, at which point the loop terminates.

\subsection{State Management and Resilience}

Session state is persisted in PostgreSQL (via asyncpg for asynchronous access), with Redis serving as a caching layer and messaging backbone. Redis Streams are used for event distribution, enabling horizontal scaling of the backend in multi-tenant deployments. Session memory uses Redis with a 24-hour TTL, with PostgreSQL as a fallback when Redis entries expire. WebSocket sessions are dropped after 20 minutes of inactivity to conserve server resources. Importantly, the system supports task reconnection: if a client disconnects due to network instability or inactivity, it can resume an in-progress task without losing prior context---a feature that proved essential for long-running coding tasks in enterprise environments where network interruptions are common.


The backend maintains a durable event timeline for every task execution, recording each tool invocation, model response, and state transition. This audit trail serves four purposes: debugging failed agent runs, providing analytics on agent behavior and tool usage patterns, satisfying compliance requirements in regulated enterprise environments, and enabling iterative self-improvement of the system. By analyzing execution logs, the team can identify recurring tool invocation errors that suggest unclear parameter schemas, detect inefficient workflows where the agent makes redundant tool calls, and observe behavioral patterns that inform refinements to tool descriptions and system prompts--a feedback loop that has proven more effective for improving agent reliability than prompt tuning in isolation.

\subsection{Tool Manifest}

A central design decision is that \textbf{tool design quality is a first-order determinant of agent effectiveness}, often more impactful than prompt engineering alone. \codegen exposes its set of tools to the LLM through a structured tool manifest---a declarative specification of each tool's name, description, parameter schema, and expected behavior. Core tools include:

\begin{itemize}[nosep]
  \item \textbf{read}: retrieves file contents and enforces a read-before-edit policy to prevent stale-context errors.
  \item \textbf{edit}: performs targeted string replacement rather than full-file rewriting, to mitigate LLM truncation failures when modifying large files.
  \item \textbf{shell}: executes terminal commands subject to multiple guardrail layers (command-level blocking, human approval mode, full audit logging).
\end{itemize}

All tool invocations are executed client-side on the developer's machine, ensuring that the agent operates on the actual project state rather than a server-side copy. 
Two execution modes--approval mode and autonomous mode--allow teams to calibrate agent autonomy as trust develops. In approval mode, only potentially destructive tools (edit and shell) require explicit human confirmation before execution; read-only operations such as read file proceed without approval, since they carry no risk of modifying the environment and requiring confirmation for every file read would introduce prohibitive friction in practice.



\section{Design Decisions}

We group our design decisions in terms of (1) Architecture and Framework Decisions (Section~\ref{sec:architecture}), (2) Tool Design and Safety (Section~\ref{sec:tool-design}), and (3) Human Oversight and Adoption (Section~\ref{sec:human-oversight}). 

\subsection{Architecture and Framework Decisions}\label{sec:architecture}

We start by describing seven architectural decisions.

\paragraph{LangChain's unidirectional chain model was inadequate for the cyclical, stateful interaction pattern required by an agentic loop.}
At project inception, the team experimented with LangChain\footnote{\url{https://www.langchain.com/}} as an orchestration framework. However, LangChain's early abstractions were designed around linear chains of components---a unidirectional pipeline where each step feeds into the next. The agentic coding assistant required a cyclical interaction pattern in which the model requests a tool, the client executes it, and the result is fed back into the model repeatedly until the task is complete. This fundamental mismatch made it difficult to express the iterative loop within LangChain's original API, \textbf{forcing the team to work around the framework rather than with it}. Framework adoption must be evaluated against the specific interaction pattern of the target system, not just its general popularity or feature set.

\paragraph{Manual implementation of the agentic loop provided more operational control and faster iteration than early framework adoption.}
After abandoning LangChain, the team implemented the agentic loop directly in project code---the Maestro component described in Section~\ref{sec:architecture}. This decision, while more labor-intensive upfront, gave the team explicit control over stop criteria, tool dispatch, WebSocket communication, and error propagation. Debugging was easier because there were no framework abstractions obscuring the execution flow. For novel or poorly understood interaction patterns, manual implementation can accelerate learning and provide clearer ownership of execution semantics---advantages that outweigh the convenience of a framework during early project phases.

\paragraph{The team is now transitioning toward modern orchestration abstractions because the framework ecosystem has converged on what was already built manually.}
As the agentic pattern matured and became widespread, frameworks evolved to support it natively. For isntance, LangChain introduced LangGraph\footnote{\url{https://www.langchain.com/langgraph}}, offering first-class support for cyclical tool-calling loops. When the team evaluated these newer abstractions, they found that the design closely resembled what had already been built by hand. This convergence validated the original architectural choices and made the transition cost low. \textbf{Building manually first and adopting frameworks later---once they mature to match actual requirements---can be a pragmatic strategy} that avoids both premature abstraction and long-term maintenance burden.

\paragraph{FastAPI was chosen for the backend due to its native async support.}
The backend must (1) sustain concurrent WebSocket connections with multiple executor clients, (2) serve SSE streams to web portals, and (3) perform asynchronous database operations---all within a single service. FastAPI's async-first design, combined with non-blocking PostgreSQL access (through asyncpg) and native WebSocket support, \textbf{provided the concurrency model needed without introducing the complexity of a multi-process architecture}. The choice of Redis Streams as the messaging backbone further supports horizontal scaling: as the number of concurrent agent sessions grows, additional backend instances can consume from the same stream group, enabling multi-tenant deployments without architectural changes.

\paragraph{Session-scoped memory already delivers substantial UX value even without long-term cross-session memory.}
Although cross-session long-term memory remains an area of ongoing development, the session-scoped approach---Redis with a 24-hour TTL, PostgreSQL fallback, and task reconnection support---already provides meaningful conversational continuity within a working session. The agent can recall prior context, tool results, and user instructions throughout an iterative task, and developers can resume interrupted sessions without starting over. A bounded, operationally simple memory layer can deliver significant user experience improvements without requiring the full complexity of persistent, long-term memory systems.


\paragraph{Internal reasoning is delegated to the LLM rather than being hand-coded in the orchestrator.}
Rather than implementing a custom reasoning engine within the Maestro, the team chose to rely on the LLM's native reasoning capabilities, controlled through a thinking effort parameter (low, medium, high) exposed by LLM APIs. The orchestrator provides the structural scaffold---context assembly, tool dispatch, output re-injection---while the model performs the cognitive work of deciding what to do next. This approach simplifies the orchestration code and accelerates product iteration, but offers less deterministic control over how the model reasons internally. With sufficiently capable models, investing in orchestration infrastructure around the model can be more productive than attempting to replicate or constrain its reasoning process.

\paragraph{Strong model capabilities reduce implementation burden but do not eliminate the need for orchestration to enforce safety boundaries.}
While delegating reasoning to the LLM simplified much of the implementation details, model capability alone does not guarantee safe or predictable behavior. The orchestrator remains essential for enforcing stop criteria, managing tool execution order, handling errors, and applying guardrails. \textbf{A powerful model can reduce the amount of hand-coded logic needed for task decomposition and planning, but it cannot replace the infrastructure that governs when and how actions are taken}. Model and orchestration are complementary dimensions: improvements in one do not substitute for investment in the other.

\subsection{Tool Design and Safety}\label{sec:tool-design}

Next we provide five decisions regarding tool design and usage.

\paragraph{Tool design quality proved more impactful on agent reliability than prompt-only tuning.}
Over the course of development, the team observed that refining how tools are described and parameterized produced more consistent improvements in agent behavior than adjusting prompts alone. The semantic quality of a tool's description determines how well the LLM understands when and why to use it. The parameter schema determines whether the model can invoke the tool correctly. And the error signaling behavior determines whether the model can recover from failed invocations. Together, these variables define the interface contract between the model and the external environment~\citep{schick2023toolformer, qin2024toolllm}. In agentic systems, \textbf{tool specification is a first-class engineering concern}---not an afterthought---and should receive the same design rigor as API design.

\paragraph{The \texttt{edit} tool was designed for targeted string replacement because LLMs tend to truncate or omit content in long file rewrites.}
When tasked with rewriting an entire file, LLMs frequently produce outputs that are incomplete, truncated, or subtly different from the intended result~\citep{xu2022systematic}. To mitigate this, the \texttt{edit} tool was designed around a targeted replacement model: the LLM specifies the file name, the old string to be found, and the new string to replace it. This constrains the model to making small, localized changes rather than regenerating entire files, significantly reducing the surface area for errors. Tool design can actively compensate for known LLM weaknesses by narrowing the scope of each individual action.

\paragraph{A read-before-edit policy prevents hallucinated edits on files the model has not recently inspected.}
Without explicit guidance, the LLM may attempt to edit files based on stale or imagined content---producing edits that reference code that no longer exists (or never existed). To address this, the system prompt instructs the model to always invoke the \texttt{read~file} tool before calling \texttt{edit}, ensuring that the model operates on the current state of the file. This policy is enforced at the prompt level rather than at the tool level, meaning it relies on model compliance rather than hard constraints. Prompt-level policies can effectively mitigate hallucination risks in tool use, though they introduce a dependency on the model's instruction-following reliability.

\paragraph{The \texttt{shell} tool is simultaneously the most useful and the most dangerous tool, requiring multiple guardrail layers.}
The \texttt{shell} tool enables the agent to run tests, invoke build tools, execute \texttt{git} commands, and perform a wide range of environment operations---making it indispensable for realistic coding workflows. However, it also grants the agent the ability to execute arbitrary commands on the developer's machine, including destructive ones such as \texttt{rm~-rf} or unauthorized \texttt{git~push}. To manage this risk, the team implemented a layered guardrail system: the tool can be blocked entirely, specific commands can be blocked granularly via a configuration file, and an approval mode can require human confirmation for every shell invocation. Additionally, all \texttt{shell} executions are recorded in the durable event timeline, providing a complete audit trail for post-hoc review. \textbf{High-utility tools with broad execution capabilities require proportionally sophisticated safety mechanisms; a single-layer restriction is insufficient}.

\paragraph{Policy consistency across tools is required because blocking one tool category is insufficient if equivalent capabilities remain accessible through another.}
During development, the team discovered that restricting a specific tool (e.g., blocking direct file deletion) could be bypassed if the \texttt{shell} tool remained unrestricted, since a shell command can achieve the same effect. Safety policies had to be designed holistically across the entire tool manifest, not on a per-tool basis. Any tool with overlapping capabilities must be subject to consistent restrictions, or the guardrails become effectively meaningless. Safety in agentic systems is a system-level property: it cannot be achieved by hardening individual tools in isolation, but requires coherent policy enforcement across all execution channels.


\subsection{Human Oversight and Adoption}\label{sec:human-oversight}

Finally, we provide four human oversights. 

\paragraph{Human approval mode serves as an effective trust-calibration mechanism during onboarding.}
\codegen offers two execution modes: approval mode, where every file edit and shell command requires explicit human confirmation, and autonomous mode, where the agent operates with minimal interruption. The team observed a consistent adoption pattern: developers begin in approval mode to understand the agent's behavior and verify its decisions, then progressively switch to autonomous mode as their confidence grows~\citep{lee2004trust}. This transition is organic and self-paced, not enforced by the agentic system. We noted that \textbf{providing a low-risk entry point is critical for adoption in enterprise environments}---human oversight mechanisms function not as permanent constraints but as transitional scaffolds that developers shed at their own pace.

\paragraph{Separating planning from execution addresses a recurrent limitation of single-pass execution.}
In the default execution flow, the LLM begins acting immediately upon receiving a prompt---reading files, editing code, running commands---without first presenting a plan for the user to approve. This single-pass behavior proved problematic for complex or high-stakes tasks, where developers wanted to verify the agent's strategy before any modifications were made~\citep{parasuraman2000model}. In response, the team developed a dedicated planning mode that generates an explicit action plan as a separate step, allowing the user to review, modify, or reject it before execution begins. Decoupling reasoning from action introduces a natural checkpoint for human oversight, improving both controllability and user confidence in the agent's behavior.

\paragraph{Progressive deployment across environments mirrors the trust-building pattern observed in individual adoption.}
Just as individual developers transition from approval mode to autonomous mode, the deployment pipeline enforces a progressive validation path: changes flow from development to staging to production through automated CI/CD gates. This organizational-level trust calibration ensures that updates to the agent's behavior---new tools, modified system prompts, updated guardrail configurations---are validated in controlled environments before reaching the full user base. The parallel between individual trust-building (approval to autonomous mode) and organizational trust-building (dev to staging to production) suggests that progressive exposure is a general principle for agent adoption, applicable at multiple levels of scale.

\paragraph{Most design decisions involved balancing competing concerns rather than optimizing a single metric.}
Throughout \codegen's development, the team repeatedly faced decisions where improving one dimension came at the cost of another. Increasing safety through approval modes added latency and friction. Delegating reasoning to the LLM simplified the codebase but reduced deterministic control. Session memory improved user experience but introduced infrastructure complexity. Choosing FastAPI and asyncpg optimized for concurrency but added complexity compared to synchronous alternatives. No single metric could serve as the optimization target; instead, each decision required explicit trade-off analysis in context. Practical \textbf{agent engineering in enterprise settings is fundamentally an exercise in trade-off management}, and the most consequential design skill is the ability to identify which trade-off matters most in each specific situation.

\section{Open Questions}

For researchers and tools builders, our learnings lead us to a few questions.

\textbf{1. How should tool manifests be designed to minimize model misuse and maximize correct invocation?} Tool descriptions, parameter schemas, and error contracts proved more impactful than prompt tuning on agent reliability. Yet there is no established methodology for designing tool specifications---teams rely on trial and error. What principles, metrics, or evaluation frameworks can guide the systematic design of tool interfaces for LLM-based agents?

\textbf{2. What is the optimal boundary between model-delegated reasoning and orchestrator-enforced control?} \codegen delegates nearly all reasoning to the LLM while the orchestrator handles safety and control flow. But this boundary was drawn pragmatically, not systematically. Under what conditions should reasoning responsibility shift from the model to the orchestrator (or vice versa), and how does this boundary affect reliability, latency, and safety as model capabilities evolve?

\textbf{3. How can safety policies be specified and enforced consistently across tools with overlapping capabilities?} The experience with the \texttt{shell} tool showed that per-tool guardrails are insufficient when multiple tools can achieve the same effect~\citep{yang2024sweagent}. What formalisms or policy languages can express cross-tool safety constraints in a way that is complete, verifiable, and maintainable as the tool manifest grows?

\textbf{4. What mechanisms best support the transition from human-supervised to autonomous agent operation?} Developers naturally migrated from approval mode to autonomous mode, but this transition was unstructured and self-paced. Are there adaptive trust models---based on task complexity, historical success rate, or action reversibility~\citep{lee2004trust, parasuraman2000model}---that can dynamically calibrate the level of human oversight required, rather than relying on a binary mode switch?

\textbf{5. How should agent memory be structured to support long-term learning without compromising session reliability?} Session-scoped memory delivered immediate UX value, but cross-session memory remains unsolved. What memory architectures balance persistence, retrieval accuracy, and staleness management across sessions---and how should agents decide what to retain, forget, or update over time without explicit user instruction?

\textbf{6. How should agent-generated code be integrated into existing quality assurance pipelines?} \codegen's experience shows that pre-commit hooks, static analysis, and test coverage requirements serve as effective guardrails for agent output. But as agents take on more complex tasks---refactoring across multiple files, modifying CI configurations, or generating migration scripts~\citep{zhang2024autocoderover}---existing quality gates may be insufficient. What additional verification mechanisms are needed, and how should they differ from those applied to human-authored code?


\section{Conclusion}

CodeGen evolved through a pragmatic, operations-first strategy rather than a framework-first strategy. The team's most consequential decisions were not limited to model selection; they centered on how reasoning is operationalized, how tools are specified and constrained, and how state is managed across interactive sessions.

The architecture---a CLI executor communicating via WebSocket with a FastAPI backend and a centralized Maestro orchestrator, backed by PostgreSQL and Redis---was shaped by the need to keep secrets and orchestration logic server-side while enabling client-side tool execution on the developer's actual environment. The choice to implement the agentic loop manually before adopting frameworks gave the team the understanding necessary to evaluate later abstractions on their merits rather than their marketing. The investment in tool design---particularly the targeted edit model, read-before-edit policy, and layered shell guardrails---produced more reliable agent behavior than prompt engineering alone. And the combination of approval mode, planning mode, and progressive deployment created an adoption path that allowed developers and the organization to build trust incrementally.

Several open questions remain, particularly around systematic tool manifest design, cross-tool safety policy enforcement, adaptive trust calibration, and long-term agent memory. These challenges are not unique to \codegen; they reflect broader gaps in the emerging practice of enterprise agent engineering. We hope that the concrete decisions and trade-offs reported here contribute to a shared understanding of what it takes to move coding agents from prototypes to production.

\bibliography{biblio}

@inproceedings{yao2023react,
  title     = {{ReAct}: Synergizing Reasoning and Acting in Language Models},
  author    = {Yao, Shunyu and Zhao, Jeffrey and Yu, Dian and Du, Nan and Shafran, Izhak and Narasimhan, Karthik and Cao, Yuan},
  booktitle = {Proceedings of the International Conference on Learning Representations (ICLR)},
  year      = {2023},
  doi       = {10.48550/arXiv.2210.03629}
}

@article{chen2021evaluating,
  title   = {Evaluating Large Language Models Trained on Code},
  author  = {Chen, Mark and Tworek, Jerry and Jun, Heewoo and Yuan, Qiming and Pinto, Henrique Ponde de Oliveira and Kaplan, Jared and Edwards, Harri and Burda, Yuri and Joseph, Nicholas and Brockman, Greg and others},
  journal = {arXiv preprint arXiv:2107.03374},
  year    = {2021},
  doi     = {10.48550/arXiv.2107.03374}
}

@inproceedings{jimenez2024swebench,
  title     = {{SWE}-bench: Can Language Models Resolve Real-World {GitHub} Issues?},
  author    = {Jimenez, Carlos E and Yang, John and Wettig, Alexander and Yao, Shunyu and Pei, Kexin and Press, Ofir and Narasimhan, Karthik},
  booktitle = {Proceedings of the International Conference on Learning Representations (ICLR)},
  year      = {2024},
  doi       = {10.48550/arXiv.2310.06770}
}

@inproceedings{yang2024sweagent,
  title     = {{SWE}-agent: Agent-Computer Interfaces Enable Automated Software Engineering},
  author    = {Yang, John and Jimenez, Carlos E and Wettig, Alexander and Liber, Kilian and Yao, Shunyu and Pei, Kexin and Press, Ofir and Narasimhan, Karthik},
  booktitle = {Advances in Neural Information Processing Systems (NeurIPS)},
  year      = {2024},
  doi       = {10.48550/arXiv.2405.15793}
}

@inproceedings{schick2023toolformer,
  title     = {Toolformer: Language Models Can Teach Themselves to Use Tools},
  author    = {Schick, Timo and Dwivedi-Yu, Jane and Dess{\`i}, Roberto and Raileanu, Roberta and Lomeli, Maria and Hambro, Eric and Zettlemoyer, Luke and Cancedda, Nicola and Scialom, Thomas},
  booktitle = {Advances in Neural Information Processing Systems (NeurIPS)},
  year      = {2023},
  doi       = {10.48550/arXiv.2302.04761}
}

@inproceedings{qin2024toolllm,
  title     = {{ToolLLM}: Facilitating Large Language Models to Master 16000+ Real-World {APIs}},
  author    = {Qin, Yujia and Liang, Shihao and Ye, Yining and Zhu, Kunlun and Yan, Lan and Lu, Yaxi and Lin, Yankai and Cong, Xin and Tang, Xiangru and Qian, Bill and others},
  booktitle = {Proceedings of the International Conference on Learning Representations (ICLR)},
  year      = {2024},
  doi       = {10.48550/arXiv.2307.16789}
}

@inproceedings{yao2023tree,
  title     = {Tree of Thoughts: Deliberate Problem Solving with Large Language Models},
  author    = {Yao, Shunyu and Yu, Dian and Zhao, Jeffrey and Shafran, Izhak and Griffiths, Tom and Cao, Yuan and Narasimhan, Karthik},
  booktitle = {Advances in Neural Information Processing Systems (NeurIPS)},
  year      = {2023},
  doi       = {10.52202/075280-0517}
}

@article{wang2024survey,
  title   = {A Survey on Large Language Model Based Autonomous Agents},
  author  = {Wang, Lei and Ma, Chen and Feng, Xueyang and Zhang, Zeyu and Yang, Hao and Zhang, Jingsen and Chen, Zhiyuan and Tang, Jiakai and Chen, Xu and Lin, Yankai and others},
  journal = {Frontiers of Computer Science},
  volume  = {18},
  number  = {6},
  pages   = {186345},
  year    = {2024},
  doi     = {10.1007/s11704-024-40231-1}
}

@article{peng2023impact,
  title   = {The Impact of {AI} on Developer Productivity: Evidence from {GitHub Copilot}},
  author  = {Peng, Sida and Kalliamvakou, Eirini and Cihon, Peter and Demirer, Mert},
  journal = {arXiv preprint arXiv:2302.06590},
  year    = {2023},
  doi     = {10.48550/arXiv.2302.06590}
}

@article{parasuraman2000model,
  title   = {A Model for Types and Levels of Human Interaction with Automation},
  author  = {Parasuraman, Raja and Sheridan, Thomas B and Wickens, Christopher D},
  journal = {IEEE Transactions on Systems, Man, and Cybernetics---Part A: Systems and Humans},
  volume  = {30},
  number  = {3},
  pages   = {286--297},
  year    = {2000},
  doi     = {10.1109/3468.844354}
}

@article{lee2004trust,
  title   = {Trust in Automation: Designing for Appropriate Reliance},
  author  = {Lee, John D and See, Katrina A},
  journal = {Human Factors},
  volume  = {46},
  number  = {1},
  pages   = {50--80},
  year    = {2004},
  doi     = {10.1518/hfes.46.1.50.30392}
}

@inproceedings{fan2023large,
  title     = {Large Language Models for Software Engineering: Survey and Open Problems},
  author    = {Fan, Angela and Gokkaya, Beliz and Harman, Mark and Lyubarskiy, Mitya and Sengupta, Shubho and Yoo, Shin and Zhang, Jie M},
  booktitle = {Proceedings of the International Conference on Software Engineering: Future of Software Engineering (ICSE-FoSE)},
  year      = {2023},
  doi       = {10.1109/icse-fose59343.2023.00008}
}

@inproceedings{zhang2024autocoderover,
  title     = {{AutoCodeRover}: Autonomous Program Improvement},
  author    = {Zhang, Yuntong and Ruan, Haifeng and Fan, Zhiyu and Roychoudhury, Abhik},
  booktitle = {Proceedings of the International Symposium on Software Testing and Analysis (ISSTA)},
  year      = {2024},
  doi       = {10.1145/3650212.3680384}
}

@inproceedings{stackspotai:pinto,
  author       = {Gustavo Pinto and
                  Cleidson R. B. de Souza and
                  Jo{\~{a}}o Batista Neto and
                  Alberto de Souza and
                  Tarc{\'{\i}}sio Gotto and
                  Edward Monteiro},
  title        = {Lessons from Building StackSpot {AI:} {A} Contextualized {AI} Coding
                  Assistant},
  booktitle    = {Proceedings of the 46th International Conference on Software Engineering:
                  Software Engineering in Practice, {ICSE-SEIP} 2024, Lisbon, Portugal,
                  April 14-20, 2024},
  pages        = {408--417},
  publisher    = {{ACM}},
  year         = {2024},
  url          = {https://doi.org/10.1145/3639477.3639751},
  doi          = {10.1145/3639477.3639751},
  bibsource    = {dblp computer science bibliography, https://dblp.org}
}

@inproceedings{xu2022systematic,
  title     = {A Systematic Evaluation of Large Language Models of Code},
  author    = {Xu, Frank F and Alon, Uri and Neubig, Graham and Hellendoorn, Vincent Josua},
  booktitle = {Proceedings of the International Symposium on Machine Programming (MAPS)},
  year      = {2022},
  doi       = {10.1145/3520312.3534862}
}

@article{rasheed2024codepori,
  title   = {{CodePori}: Large-Scale System for Autonomous Software Development Using Multi-Agent Technology},
  author  = {Rasheed, Zeeshan and Sami, Muhammad and Waseem, Muhammad and Kemell, Kai-Kristian and Wang, Xiaofeng and Duc, Anh Nguyen and Abrahamsson, Pekka},
  journal = {SSRN Electronic Journal},
  year    = {2024},
  doi     = {10.2139/ssrn.4979510}
}

\end{document}